\input epsf

\documentstyle{amsppt}
\magnification\magstep1

\NoBlackBoxes

\baselineskip=12pt

\hsize=31pc
\hoffset=4pt
\parindent=1.5em
\pageheight{45.5pc}

%%%%%%%%%%%%%%%%%%%%%%%%%%%%%%%%%%%%%%%%%%%%%%%%%%%%
%            DEFINITIONS                           %
%%%%%%%%%%%%%%%%%%%%%%%%%%%%%%%%%%%%%%%%%%%%%%%%%%%%

\define\1{\'{\i}}                           
\define\k{{\kappa}}
\define\diag{{\text{\,diag\,}}}

\define\Sk{\text{\ \!S}}            
\define\Ck{\text{\ \!C}}           
\define\Tk{\text{\ \!T}}            
\define\Vk{\text{\ \!V}}

\define\dd{\text {d}}
\define\>#1{{\bold#1}}                 
\define\te{\theta}
\define\a{a}

 \define\pp{\pi}
\define\be{\beta}
\define\dis{x}

\define\inte{Q}

\define\tea{r}

\define\tildep{{\tilde P}}
\define\tildej{{\tilde J}}

\font\head=cmbx12
\define\endhead{\rm}
\font\subhead=cmbx10
\define\endsubhead{\rm}
\font\titulo=cmbx10 scaled\magstep2

%%%%%%%%%%%%%%%%%%%%%%%%%%%%%%%%%%%%%%%%%%%%%%%%%%%%

%%%%%%%%%%%%%%%%%%%%%%%%%%%%%%%%%%%%%%%%%%%%%%%%%%%%

\ 
\vskip 1truecm

\centerline {\titulo Maximally  superintegrable   
Smorodinsky-Winternitz systems }
\bigskip
\centerline {\titulo on the N-dimensional sphere and hyperbolic
spaces\footnote{Published in {\it Superintegrability in  Classical and Quantum  
Systems}, edited by P.~Tempesta, P.~Winternitz,   J.~Harnad, W.~Miller Jr., G.~Pogosyan
and M.A.~Rodr\1guez,  CRM Proceedings \& Lecture Notes, vol.~37, American Mathematical
Society, 2004 }}
\bigskip

\vskip 1truecm
\bigskip

\centerline {Francisco J. Herranz$^a$, Angel Ballesteros$^a$, Mariano Santander$^{b}$,
and Teresa Sanz-Gil$^{b}$}
\bigskip
\bigskip

\centerline{\it { 
${}^a$Departamento de F\1sica, Facultad de Ciencias, Universidad de
Burgos,   }}

\centerline{\it {  E-09001 Burgos, Spain }}

\bigskip

\smallskip 
\centerline{\it { 
${}^b$Departamento de F\1sica Te\'orica, Facultad de Ciencias, 
Universidad de Valladolid,   }}

\centerline{\it E-47011 Valladolid, Spain}
\smallskip

\bigskip
\bigskip
\bigskip
\bigskip

\noindent {\bf Abstract.}	The classical   Smorodinsky--Winternitz
systems  on the $N$D sphere, Euclidean and hyperbolic spaces $\Bbb S^N$,
$\Bbb E^N$ and  $\Bbb H^N$ 
are simultaneously approached starting from the
Lie algebras $\frak{so}_\k(N+1)$, which include a parametric dependence on the
curvature. General expressions for the Hamiltonian and its integrals of motion are
given in  terms of intrinsic geodesic coordinate systems.    Each Lie algebra
generator gives rise to an integral of motion, so that a set of $N(N+1)/2$
integrals  is   obtained. Furthermore,  $2N-1$ functionally independent ones are
identified which, in turn, shows that  the well known maximal superintegrability  
of  the  Smorodinsky--Winternitz system on $\Bbb E^N$ is preserved when curvature
arises.  On both $\Bbb S^N$ and $\Bbb H^N$, the resulting system can be interpreted
as a superposition of an ``actual" oscillator and $N$ ``ideal" oscillators (for
the sphere, these are alike the actual ones),   which can also be  understood as
$N$ ``centrifugal terms"; this is the form seen in the Euclidean limiting case.

\newpage

\document

%%%%%%%%%%%%%%%%%%%%%%%%%%%%%%%%%%%%%%%%%%%%%%%%%%%%

\noindent
\head 1. Introduction\endhead

\medskip
\noindent
Superintegrable systems on the two- and three-dimensional
(3D) Euclidean spaces have been classified in~\cite{6, 25}, 
and also extended to   the 2D   and 3D
sphe\-res~\cite{11} as well as to the 2D  hyperbolic
plane~\cite{19, 20} and 3D hyperbolic space~\cite{12}.
Recent   classifications of superintegrable systems for these 2D
Riemannian spaces have been presented  in~\cite{18, 21, 24, 27}. 
In particular,  in the 2D sphere $\Bbb S^2$ there are two (maximally)
superintegrable families: the harmonic oscillator  and the
Kepler--Coulomb potential, both of them with some ``additional" terms. Let
us   consider $\Bbb S^2$ as embedded through $s_0^2+s_1^2+s_2^2=1$  in an
ambient space $\Bbb R^3=(s_0,s_1,s_2)$;   we set the  geodesic
polar coordinates
$(\tea,\te)$   such that 
$ s_0=\cos\tea$, $s_1=\sin\tea\cos\te$, 
$s_2=\sin\tea \sin\te$. 
Following the notation and results given
in~\cite{27},  the first classical superintegrable family on $\Bbb
S^2$ is given by
$$
\aligned
{\Cal U}_{\text{ho}}&=  \be_0\left(\frac {   s_1^2+s_2^2}{s_0^2}\right)+  
\frac {\be_1}{s_1^2} +\frac {\be_2}{s_2^2} \\
&=\be_0
\tan^2 \tea + \frac{\be_1}{\sin^2\tea\cos^2\te}
  +\frac{\be_2}{ \sin^2\tea \sin^2\te} ,
\endaligned
\tag1.1
$$
where $\be_i$ are real constants. The first term, $\tan^2\tea,$ is the
{\it spherical oscillator} or Higgs potential~\cite{16, 26}. Under contraction
to the Euclidean plane, this reduces to the usual harmonic oscillator,
$r^2$, while the two remaining  terms  give rise to two ``centrifugal
barriers". However, very recently, all the three terms in ${\Cal
U}_{\text{ho}}$ have been interpreted as spherical oscillators with
different  centers~\cite{29}.

On the other hand, the second superintegrable family on $\Bbb
S^2$ turns out to be  
$$
\aligned
{\Cal U}_{\text{KC}}&=  \be_0\,\frac { s_0}{\sqrt{ s_1^2+s_2^2}} +  
\be_1\,\frac {s_1}{s_2^2\sqrt{ s_1^2+s_2^2}} +\frac {\be_2}{s_2^2} \\
&=
\be_0\,\frac{1}{\tan\tea} + \be_1\,\frac{\cos\te }{\sin^2\tea\sin^2\te}
  + \frac{\be_2}{ \sin^2\tea \sin^2\te} ,
\endaligned
\tag1.2
$$
where $1/\tan r$ is the ``spherical" Kepler--Coulomb potential,
first studied by Schr\"o\-dinger~\cite{30}; the two potentials 
$\tan^2 r$ and
$1/\tan r$ are mutually related~\cite{22, 23}.

The aim of this contribution is to study the maximal
superintegrability of the generalization
of the potential (1.1)   on   the
$N$D spaces $\Bbb S^N$,
$\Bbb E^N$ and   $\Bbb H^N$   from
a group theoretical standpoint. 
% , which may be also applied to the second
% potential (1.2).  
This family, depending on the curvature $\k$ as a parameter, includes for
$\k=0$ the well known maximally superintegrable Euclidean 
Smorodinsky--Winternitz (SW) system~\cite{7--10}, the Hamiltonian of
which reads
$$
 {\Cal H}=\frac12\sum_{i=1}^N\left( p_i^2+ 2\be_0
q_i^2+ \frac{2\be_i}{q_i^2}\right) ,
\tag1.3
$$
where $\sum_iq_i^2\equiv r^2$ is the harmonic oscillator potential and each
$1/{q_i^2}$ is a ``centrifugal term".  Two sets of  
integrals of motion  for $ {\Cal H}$ are given by $(i<j;\ i,j=1,\dots,N)$:
$$
\alignedat2
&  I_{0i}=\tildep_i^2+2\be_0
q_i^2+2\be_i\frac{1}{q_i^2},&\quad \text{with}\quad   &\tildep_i=p_i,\\ 
& I_{ij}=\tildej_{ij}^2+2\be_i
\frac{q_j^2}{q_i^2}+ 2\be_j
\frac{q_i^2}{q_j^2},&\quad \text{with}\quad 
&\tildej_{ij}=q_ip_j-q_jp_i  .
\endalignedat
\tag1.4
$$
  The first  set  comes  from the separability of the Hamiltonian
$2{\Cal H}=\sum_i I_{0i}$, while the second one has a term quadratic in
momenta (the square of the components of the angular momentum
tensor) plus some additional terms depending on coordinates.  The
functions
$\tildep_i$,
$\tildej_{ij}$ close in phase space the commutation relations of the
Euclidean algebra
$\frak{iso}(N)$  with respect to the   canonical Lie--Poisson bracket:
$$
\{f,g\}=\sum_{i=1}^N\left(\frac{\partial f}{\partial q_i}
\frac{\partial g}{\partial p_i}
-\frac{\partial g}{\partial q_i} 
\frac{\partial f}{\partial p_i}\right)  .
\tag1.5
$$

The structure of this contribution is as follows. The next section
contains the details required (related to their maximal groups of
isometries) on the  three $N$D classical Riemannian spaces with constant
curvature $\k$. Geodesic motion is obtained in section 3
starting from the metric on these spaces; this is the kinetic energy
term to which possible potentials can be  added. 
The   generalization of the SW family to these spaces (any $\k$) is
performed in section 4. General expressions for the Hamiltonian and its
integrals of motion are explicitly given both in terms of the 
(Weierstrass) coordinates in an ambient linear  space $\Bbb R^{N+1}$
as well as by means of two sets of intrinsic coordinates, 
the non-zero curvature versions of the Euclidean Cartesian and polar
ones. Moreover, $2N-1$ integrals, including the Hamiltonian, are shown to
be functionally independent, thus proving that the SW system on curved
spaces is also maximally superintegrable \cite{4}. As an example, we apply
in the section 5 the general expressions to the 
$N=4$ case.

%%%%%%%%%%%%%%%%%%%%%%%%%%%%%%%%%%%%%%%%%%%%%%%%%%%%

\bigskip
\bigskip
\noindent
\head 2. The sphere $\Bbb S^N$, Euclidean $\Bbb E^N$, and hyperbolic
$\Bbb H^N$ spaces\endhead

\medskip
\noindent
Let $\frak{so}_\k(N+1)$ be the Lie algebra of the motion group  $SO_\k(N+1)$
on a generic $N$D real Riemannian  space with constant
curvature $\k$, denoted $S^N_{[\k]}$. In the basis $\{J_{0i}\equiv
P_i,J_{ij}\}$ ($i,j=1,\dots,N$; $i<j$), the non-vanishing commutation
relations of   $\frak{so}_\k(N+1)$ are given by 
$$
\alignedat3
&[J_{ij},J_{ik}]=J_{jk},&\qquad &[J_{ij},J_{jk}]=-J_{ik},&\qquad 
&[J_{ik},J_{jk}]=J_{ij},\\
&[J_{ij},P_{i}]=P_j,&\qquad &[J_{ij},P_{j}]=-P_i, &\qquad &[P_i,P_j]=\k
J_{ij}, 
\endalignedat
\tag2.1
$$
with $i<j<k$. The Lie algebra  $\frak{so}_\k(N+1)$ is isomorphic to either
  $\frak {so}(N+1)$ for $\k>0$,   $\frak {iso}(N)$ for $\k=0$, or   
$\frak {so}(N,1)$ for $\k<0$. Notice that  by scaling the generators $P_i$,
any value of $\k$ can always be reduced to either $+1$, 0 or $-1$. 
The involutive automorphism defined by
$$
\Theta\ :\  J_{ij}\longrightarrow J_{ij}  ,\qquad P_i\longrightarrow -P_i
,\qquad
i,j=1,\dots,N,
\tag2.2
$$
provides the following Cartan decomposition of $\frak{so}_\k(N+1)$:
$$
\frak {so}_\k(N+1)=\frak h\oplus \frak p,\qquad \frak h=\langle
J_{ij}\rangle=\frak {so}(N),\qquad
 \frak p=\langle P_{i}\rangle .
\tag2.3
$$
The generators invariant under $\Theta$ span the subalgebra  $\frak h$ 
of the Lie subgroup $H\simeq SO(N)$, so that 
$ S^N_{[\k]}=SO_\k(N+1)/SO(N)$ is a family, parametrized by
$\k$, of 
$N$D symmetric homogeneous rank-one spaces~\cite{13}. The generators
$J_{ij}$  leave an {\it origin} point $\Cal O$ invariant, thus acting as
rotations around $\Cal O$, while the remaining 
$P_i$ generate translations that move $\Cal O$ along $N$ basic geodesics
$l_i$ orthogonal at $\Cal O$. The space $ S^N_{[\k]}$    comprises the
three classical Riemannian spaces:  
$$
\alignedat3
&\k>0,&\qquad &\text{Sphere},&\qquad &S_{[+]}^N\equiv SO(N+1)/SO(N)\equiv
\Bbb S^N;\\ &\k=0,&\qquad &\text{Euclidean},&\qquad &S_{[0]}^N\equiv
ISO(N)/SO(N)\equiv
\Bbb E^N;\\
&\k<0,&\qquad &\text{Hyperbolic},&\qquad &S_{[-]}^N\equiv
SO(N,1)/SO(N)\equiv
\Bbb H^N.
\endalignedat
$$
The curvature $\k$ can also be interpreted as a graded contraction
parameter coming from the $\Bbb Z_2$-grading of $\frak {so}_\k(N+1)$
determined by $\Theta$; the value $\k=0$ corresponds to the
contraction $\Bbb S^N\rightarrow \Bbb E^N\leftarrow  \Bbb H^N$ around a
point (the origin $\Cal O$) of the spaces.

The  Killing--Cartan form  of  $\frak {so}_\k(N+1)$,  
$g^{KC}(J_{ij},J_{kl})= \text{Trace}(\text{ad}\, J_{ij}\cdot
\text{ad}\, J_{kl})$,   is diagonal   with the following
non-zero elements 
$$
g^{KC}(P_{i},P_{i})=-2(N-1)\k,\qquad  g^{KC}(J_{ij},J_{ij})=-2(N-1); 
\tag2.4
$$
hence the restriction of $g^{KC}$ to the subspace ${\frak p}$, can be written as:
$$
g^{KC}\bigl|_{\frak p}= -2(N-1)\k\,g\bigl|_{\Cal O} ,\qquad g\bigl|_{\Cal
O}(P_i,P_j) =
\delta_{ij},
\tag2.5
$$
where $g\bigl|_{\Cal O}$ is to be considered as the metric in the
tangent space
${\frak p}$ at $\Cal O$, represented by the 
$N\times N$  matrix
$\diag (+,+,\dots,+)$. This metric can be translated to all points of
$ S^N_{[\k]}$ by the group action. Even if $g^{KC}$ vanishes in the
Euclidean case, the choice $g\propto g^{KC}/\k$ ensures a non-degenerate
metric in
$S^N_{[\k]}$ $\forall\k$.

%\newpage

%%%%%%%%%%%%%%%%%%%%%%%%%%%%%%%%%%%%%%%%%%%%%%%%%%%%

\bigskip
\medskip
\noindent
\subhead 2.1. Vector model and Weierstrass coordinates\endsubhead

\medskip
\noindent
The {\it vector representation} of $\frak{so}_\k(N+1)$  is given
by  $(N+1)\times (N+1)$ real matrices  fulfilling  (2.1):
$$
P_i=-\k \, e_{0i}+e_{i0},\qquad J_{ij}=- e_{ij}+e_{ji},
\tag2.6
$$
where $e_{ij}$ is the matrix with a single non-zero entry $1$, at row $i$
and column $j$. Their exponential gives rise to the following one-parametric
subgroups of $SO_\k(N+1)$:
$$
\aligned 
\exp\{x P_i\}&=
\!\!{\sum_{s=1;s\ne i}^N}\! e_{ss}+ 
\Ck_{\k}(x)e_{00}+\Ck_{\k}(x)e_{ii}
-\k \Sk_{\k}(x)e_{0i}+\Sk_{\k}(x)e_{i0}\\
&=\Bbb I+P_{i}\Sk_{\k}(x)+ P^2_{i} \Vk_\k(x)  ,\\
\exp\{x J_{ij}\}&=
\!\!{\sum_{s=0;s\ne i,j}^N}\! e_{ss}+ 
\cos x \,e_{ii}+\cos x\, e_{jj}
-\sin x\, e_{ij}+\sin x\,e_{ji}\\
&=\Bbb I+J_{ij}\sin x+J^2_{ij}(1-\cos x) ,
\endaligned 
\tag2.7 
$$
where $\Bbb I$  is the $(N+1)\times (N+1)$ identity matrix. The   
 curvature-dependent cosine $\Ck_{\k}(x)$ and sine $\Sk_{\k}(x)$ functions
are defined by~\cite{3}:
$$ 
\aligned 
&\Ck_{\k}(x) =\sum_{l=0}^{\infty}(-\k)^l{x^{2l}\over 
(2l)!}=\left\{
\aligned
  \cos {\sqrt{\k}\, x} ,\qquad\quad  \k&>0; \\ 
\qquad 1 , \qquad\qquad
  \k&=0 ;\\ 
\cosh {\sqrt{-\k}\, x} ,\quad\ \   \k&<0 .
\endaligned
\right. \\
&  \Sk{_\k}(x) =\sum_{l=0}^{\infty}(-\k)^l{x^{2l+1}\over 
(2l+1)!}
= \left\{
\aligned
  \tfrac{1}{\sqrt{\k}} \sin {\sqrt{\k}\, x} ,\qquad\quad  \k&>0 ;\\ 
\qquad x , \qquad\qquad
  \k&=0 ;\\ 
\tfrac{1}{\sqrt{-\k}} \sinh {\sqrt{-\k}\, x} ,\quad\ \   \k&<0 .
\endaligned
\right.  
\endaligned
\tag2.8
$$
From them, we define the ``versed sine" or versine $\Vk_\k(x)$ and the
tangent $\Tk_\k(x)$:
$$
\Vk_\k(x)=\frac 1  \k (1-\Ck_\k(x)),\qquad
\Tk_\k(x)=\frac{\Sk_\k(x)}{\Ck_\k(x)} .
\tag2.9
$$
These $\k$-trigonometric functions coincide with the usual elliptic
(circular) and hyperbolic ones for $\k=+1$ and $\k=-1$, respectively. The
``flat" case with $\k=0$  gives the parabolic  functions: 
$\Ck_{0}(x)=1$, $\Sk_{0}(x)=x$ and $\Vk_{0}(x)=x^2/2$. In this sense, they
can be interpreted as the {\it ``curvature $\k$-deformed versions"} of 
$1$, $x$, and $x^2$. All the known trigonometric  relations  
(necessary in the further computations) also extend  for these
$\k$-functions~\cite{14, 15} such as, for instance,
$$
\Ck^2_\k(x)+\k\,\Sk^2_\k(x)=1,\qquad \frac{ {\text d}}
{\text{d} x}\Ck_\k(x)=-\k\,\Sk_\k(x),\qquad 
\frac{ {\text d}}
{\text{d} x}\Sk_\k(x)= \Ck_\k(x) .
$$

In the vector representation (2.6), any   generator $X$ satisfies  the
relation
$$
 X^T {\Bbb I}_\k+{\Bbb I}_\k X=0,\qquad 
{\Bbb I}_\k= e_{00}+\k\sum_{i=1}^{N}
e_{ii}=\diag(1,\k,  \dots,  \k ),
\tag2.10
$$
($X^T$ is the transpose matrix of $X$), so that    any element $R\in 
SO_\k(N+1)$ verifies $R^T {\Bbb I}_\k  R={\Bbb I}_\k$. 
In this way,  $SO_\k(N+1)$ can be seen as a group of linear 
isometries of the bilinear form 
${\Bbb I}_\k$ acting on $\Bbb R^{N+1}=(s_0,s_1,\dots,s_N)$ via matrix
multiplication. The action of $SO_\k(N+1)$ on $\Bbb R^{N+1}$ is linear
but not transitive, since it conserves the quadratic form 
$$s_0^2+\k\sum_{i=1}^N s_i^2$$ provided by ${\Bbb I}_\k$, and
$H\simeq
SO(N)=\langle J_{ij}\rangle$ (2.3) is the isotopy subgroup of the point
${\Cal O}=(1,0,\dots,0)\in \Bbb R^{N+1}$ which will be taken as the {\it
origin} in the space
 $ S^N_{[\k]}$. The action becomes transitive if we restrict to  the
orbit in $\Bbb R^{N+1}$ of the point $\Cal O$, which is contained in the
``sphere" $\Sigma$:
$$
\Sigma \ \equiv\     s_0^2+\k\sum_{i=1}^N s_i^2=1 ,
\tag2.11
$$
which reproduces the whole hypersphere, two  hyperplanes and a
two-sheeted hyperboloid for $\k>,=,<0$, respectively.
This orbit is identified with the  space $ S^N_{[\k]}$, and
$(s_0,s_1,\dots,s_N)$, fulfilling  the ``sphere" constraint (2.11), are
called {\it Weierstrass coordinates}.   
 In terms of these,  the metric on $ S^N_{[\k]}$ comes 
from the flat ambient metric in $\Bbb R^{N+1}$ divided by $\k$ and
restricted to $\Sigma$:
$$
\dd
\sigma^2=\left.{1\over\k}
\left(\dd s_0^2+\k  \sum_{i=1}^{N}\dd
s_i^2\right)\right|_{\Sigma} =
\sum_{i=1}^{N}\dd
s_i^2 + \k\, \frac{\left(\sum_{i=1}^{N} 
s_i \dd s_i\right)^2}{1-\k \sum_{i=1}^{N}\dd
s_i^2} ,
\tag2.12
$$
which reduces to the Euclidean one for $\k=0$. A  differential realization
of the  generators as first-order vector fields in $\Bbb R^{N+1}$ with
$\partial_i = {\partial}/{\partial s_i}$ is (see (2.6)):
$$
P_i=\k \, s_i\partial_0 - s_0\partial_i, \qquad 
J_{ij}=  s_j\partial_i - s_i\partial_j .
\tag2.13
$$

%%%%%%%%%%%%%%%%%%%%%%%%%%%%%%%%%%%%%%%%%%%%%%%%%%%%
\bigskip
\medskip
\noindent
\subhead 2.2. Geodesic coordinate systems\endsubhead

\medskip
\noindent
 Let us consider in
the vector model the origin
$\Cal O=(1,0,\dots,0)$ and a generic point $Q\in
S^N_{[\k]}$ with Weierstrass coordinates $\>s=(s_0,s_1,\dots,s_N)\in\Bbb
R^{N+1}$.   Starting from   $\Cal O$, the point $Q$ can be
reached in different ways through the action of $N$
one-parametric subgroups (2.7), that is, by means of motions
in the space
$S^N_{[\k]}$. In this way, we introduce coordinates which are
intrinsic quantities to the space $S^N_{[\k]}$ itself, the
associated Weierstrass coordinates automatically fulfilling the condition
(2.11). We shall consider two possibilities.

\newpage

\bigskip
\noindent
\subsubhead \noindent{\rm 2.2.1.} Geodesic parallel coordinates
$\a=(\a_1,\dots,\a_N)$\endsubsubhead

\medskip
\noindent
We move the origin $\Cal O$ by using 
 the $N$ translations
subgroups   as
$$
\aligned
 \>s(\a)
&=\exp(\a_1 P_1)\,\exp(\a_2 P_2)\dots\exp(\a_N P_N)\,  \Cal O \\
 \pmatrix s_0\\ s_1\\ s_2\\ \vdots \\ s_{N-1}\\ s_N\endpmatrix
&=\pmatrix
\Ck_{\k}(\a_1)\Ck_{\k}(\a_2)\Ck_{\k}(\a_3)\dots
\Ck_{\k}(\a_N)\\
\Sk_{\k}(\a_1)\Ck_{\k}(\a_2)\Ck_{\k}(\a_3)\dots
\Ck_{\k}(\a_N)\\
\Sk_{\k}(\a_2)\Ck_{\k}(\a_3)\dots \Ck_{\k}(\a_N)\\
 \vdots \\
\Sk_{\k}(\a_{N-1})\Ck_{\k}(\a_N)\\
 \Sk_{\k}(\a_N)
\endpmatrix .
\endaligned
\tag2.14
$$
Each coordinate  $\a_i$ is  associated to the generator $P_i$ and has
dimensions of {\it length}. If  $l_1,l_2,\dots,l_N$ are the $N$ (oriented) 
basic geodesics in $S^N_{[\k]}$  orthogonal  at $\Cal O$, then 
the first coordinate of a point $Q$ is the geodesic distance $\a_1$
between $\Cal O$ and $Q_1$ (the orthogonal projection of $Q$ on
$l_1$), measured along  
$l_1$. The second coordinate  is the distance $\a_2$ between $Q_1$ and
another point
$Q_2$, measured along a geodesic  $l'_2$ orthogonal to $l_1$
 through $Q_1$ and parallel to $l_2$ (in the sense of parallel transport)
and so on, up to reaching $Q$~\cite{15}. This is depicted in figure 1 for
$\Bbb S^{2}$ and $\Bbb H^{2}$. By introducing  (2.14)  in  (2.12),   we obtain the
  metric:
$$
\dd \sigma^2=
\sum_{i=1}^{N-1} \left( \prod_{l={i+1}}^N\!\!\Ck^2_{\k}(\a_l)\right)
\dd\a_i^2+ \dd\a_N^2 .
\tag2.15
$$
 From it we may compute the Levi-Civita connection, whose 
non-zero  Christoffel symbols   are ($i =1,\dots,N-1;\  
i+1\le j\le N$):
$$
\Gamma_{\ ij}^i=-\k \Tk_{\k}(\a_j), \qquad
 \Gamma_{\ ii}^j =\k \Tk_{\k}(\a_j)
\!\!\prod_{l=i+1}^j\!\!\Ck_{\k}^2(\a_l) .
\tag2.16
$$

\topinsert
\epsfbox{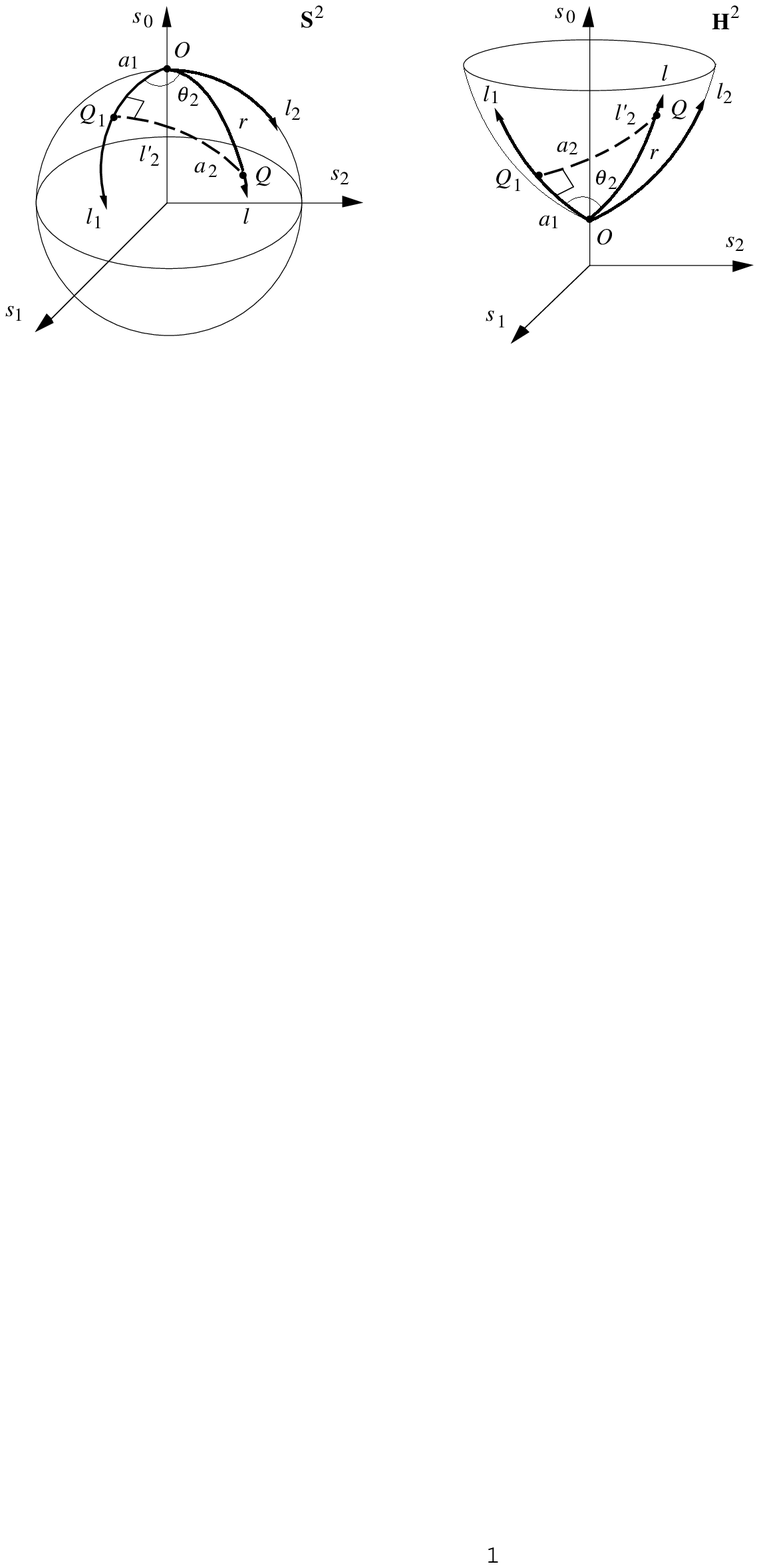}
\botcaption{Figure 1}
Geodesic parallel $(\a_1,\a_2)$ and polar $(\tea,\te_2)$ coordinates in
$\Bbb S^2$ and $\Bbb H^2$.
\endcaption
\endinsert

%\topinsert
%\vspace{5.5cm}
%\epsfbox{figure1.eps}
%\epsfbox[0 0 30 50]{figure1.eps}
%\epsfxsize=10cm\epsfbox{figure1.eps}
%\epsfysize=10cm\epsfbox{figure1.eps})  
%\ttcs{captionwidth}\{\Dimen\}\quad{\rm(optional)}
%\botcaption{Figure 1}
%Geodesic parallel $(\a_1,\a_2)$ and polar $(\tea,\te_2)$ coordinates in
%$\Bbb S^2$ and $\Bbb H^2$.
%\endcaption
%\endinsert

%%%%%%%%%%%%%%%%%%%%%%%%%%%%%%%%%%%%%%%%%%%%%%%%%%%%

\bigskip
 \noindent
\subsubhead  \noindent{\rm 2.2.2.} Geodesic polar coordinates
$\te=(\tea,\te_2,\dots,\te_N)$\endsubsubhead  

 \medskip
\noindent
In this case, we move $\Cal O$ by using the $N-1$ rotations $J_{i\,i+1}$ as
well as the first translation $J_{01}\equiv P_1$:
$$
\aligned
 \>s(\te) &=\exp(\te_N J_{N-1\, N})\dots\exp(\te_2 J_{12})\,
\exp(\tea P_{1})\, \Cal O\\
\pmatrix s_0\\ s_1\\ s_2\\ \vdots \\ s_{N-1}\\ s_N\endpmatrix
&=\pmatrix
\Ck_{\k}(\tea) \\
\Sk_{\k}(\tea)\cos\te_2  \\
\Sk_{\k}(\tea)\sin \te_2 \cos\te_3 \\
\quad  \vdots \\
\Sk_{\k }(\tea)\sin \te_2\dots\sin \te_{N-1} \cos \te_N \\
\Sk_{\k }(\tea)\sin \te_2\dots\sin\te_{N-1} \sin\te_{N} 
\endpmatrix .
\endaligned
\tag2.17
$$
The  ``radial" coordinate $\tea$ associated to $P_1$ has
dimensions of {\it length} and is the distance  between
$\Cal O$ and $Q$ measured along the geodesic $l$ joining both points.
The   remaining  $\te_2,\dots,\te_N$ are ordinary {\it angles}
parametrizing the orientation of $l$ with
respect to the reference flag at  
$\Cal O$ spanned by $\{l_1,l_2,\dots,l_{i-1}\}$ (see figure 1 for $N=2$).
On the sphere
$\Bbb S^N$ of radius $R$ and curvature $\k=1/R^2$, all the usual spherical
coordinates are  angles, and these differ from the geodesic polar
ones  only in the first coordinate~\cite{28}, which is commonly taken as the
dimensionless quantity $\te_1 \equiv r/R$. While these conventional
 spherical coordinates  require an explicit contraction   to
the Euclidean ones, as done for instance in~\cite{17}, our
choice  works for {\it any} value of $\k$ and  when
$\k=0$ we  recover  directly, without any limiting procedure,  the 
  polar (and Cartesian) coordinates in $\Bbb E^N$.

 In polar coordinates the
metric (2.12) turns out to be  
$$
\dd \sigma^2=\dd\tea^2+\Sk^2_{\k}(\tea)\left( \dd\te_2^2+
  \sum_{i=3}^{N} \,
\biggl(\,\prod_{l={2}}^{i-1} \sin^2 \te_l\biggr) \dd\te_i^2 \right).
\tag2.18
$$

The components of the Riemann and Ricci tensors can be computed in
either geodesic coordinate system; the
scalar and  the   sectional curvatures of $S^N_{[\k]}$
are   $R=N(N-1)\k$ and   $K=\k$, both constant, respectively. When $\k=0$,
the expressions (2.15) and (2.18) give   the Euclidean metric
$$
\dd \sigma^2= \sum_{i=1}^{N}  \dd\a_i^2= \dd\tea^2+ \tea^2\left(
\dd\te_2^2+ \sum_{i=3}^{N}\, 
\biggl(\,\prod_{l={2}}^{i-1} \sin^2 \te_l\biggr) \dd\te_i^2 \right),
\tag2.19
$$
and   all the Christoffel symbols $\Gamma_{\ jk}^i$ (2.16)
vanish in parallel coordinates.

\newpage

%%%%%%%%%%%%%%%%%%%%%%%%%%%%%%%%%%%%%%%%%%%%%%%%%%%%

\bigskip
\bigskip
\noindent
\head 3.  
Geodesic motion  on $S^N_{[\k]}$ and phase space realization of
$\frak {so}_\k(N+1)$\endhead

\medskip
\noindent
If we now adopt a dynamical viewpoint, the expressions of the metric in
$S^N_{[\k]}$ (2.15) and (2.18) provide  the kinetic energy $\Cal T$ of a
particle in terms of the velocities, generically denoted as $\dot
q$ (either $\dot
\a$ or $\dot\te$) in the coordinate systems (2.14) and (2.17):
$$
2{\Cal T}= \sum_{i=1}^{N-1}\,
 \!\biggl(\, \!\!\!\!\!\!\prod_{\ \ \ l=i+1}^N
\!\!\!\!\!\Ck_\k^2(\a_l)\!\biggr){\dot\a}_i^2  +  {\dot\a}_N^2 = 
\dot\tea^2+\Sk^2_{\k}(\tea)\! \left(\! \dot\te_2^2+
\sum_{i=3}^{N} \, \biggl(\,\prod_{l={2}}^{i-1} \sin^2 \te_l\biggr)
\dot\te_i^2\right)  ,
\tag3.1
$$
which is  the Lagrangian  
${\Cal L}\equiv{\Cal T}$ of the  geodesic motion on  
$S^N_{[\k]}$.  The   canonical
momenta,  $p=\partial{\Cal L}/\partial{\dot q}$, (denoted $p, \pp$ for
$q=\a,\te$) are given by:
$$
\aligned
&p_i=\left(\prod_{l=i+1}^N \!\!\Ck_\k^2(\a_l)\right){\dot\a}_i,\quad
i=1,\dots,N-1;\qquad p_N=  {\dot\a}_N,\\
&\pp_1= \dot\tea,\quad \pp_2=\Sk^2_{\k}(\tea) \dot\te_2
,\quad  \pp_j=\Sk^2_{\k}(\tea)\left(\prod_{l={2}}^{j-1} \sin^2
\te_l\right)
\dot\te_j,\ \ j=3,\dots,N.
\endaligned
\tag3.2
$$
By introducing these momenta in (3.1),  we obtain the  free Hamiltonian
${\Cal H}\equiv{\Cal T}$ in the phase space of motions in $S^N_{[\k]}$
expressed in either ``parallel" $(\a,p)$ or ``polar" $(\te,\pp)$ canonical
coordinates and momenta:
$$
\aligned
2{\Cal T}&= \sum_{i=1}^{N-1}
\frac{ p_i^2 }{\prod_{l=i+1}^N \!\!\Ck_\k^2(\a_l)} +  p_N^2= \pp_1^2+
\frac{1}{\Sk^2_{\k}(\tea)}  \left({\pp_2^2} +
\sum_{i=3}^{N}\frac{\pp_i^2}{ \prod_{l={2}}^{i-1} \sin^2
\te_l}  \right) .
\endaligned
\tag3.3
$$
 
On the other hand, the Lie generators are expressed in terms of
Weierstrass coordinates as:
$$
\tildep_{i}(s(q),\dot s(q,p))= s_0 {\dot s}_i-   s_i {\dot s}_0
,\qquad \tildej_{ij}(s(q),\dot s(q,p))= s_i {\dot s}_j-   s_j {\dot s}_i,
\tag3.4 
$$
and thus we get an $N$-particle realization of  $\frak{so}_\k(N+1)$ 
in the  phase space simply by rewriting everything either in terms of  
$(\a,p)$ or  
$(\te,\pp)$. The time derivatives of
$\>s$ are obtained from either  (2.14) in terms
of parallel coordinates and velocities
$(\a,\dot\a)$   
$(i=1,\dots,N-1)$: 
$$
\aligned
&{\dot s}_0=-\k\prod_{m=1}^N\!\Ck_\k(\a_m)\sum_{l=1}^N\Tk_\k(\a_l){\dot
\a}_l,\\
&{\dot s}_i=\prod_{m=i}^N \!\Ck_\k(\a_m)\left({\dot \a}_i-\k\Tk_\k(\a_i)
\sum_{l=i+1}^N\!\Tk_\k(\a_l){\dot \a}_l
\right) , \qquad {\dot s}_N= \Ck_\k(\a_N){\dot \a}_N ,
\endaligned
\tag3.5
$$
or from (2.17) in terms of polar coordinates
and velocities $(\te,\dot\te)$   $(j=2,\dots,N-1)$:
$$
\aligned
&{\dot s}_0=-\k\Sk_\k(\tea)\dot\tea, \qquad {\dot s}_1=
\Sk_\k(\tea)\sin\te_2\left(
\frac{ \dot\tea}{ \Tk_\k(\tea)\tan\te_2}- \dot\te_2\right),\\
&{\dot s}_j= \Sk_\k(\tea)\prod_{m=2}^{j+1}\sin\te_m\left(
\frac{ \dot\tea}{ \Tk_\k(\tea)\tan\te_{j+1}}
+\sum_{l=2}^j\frac{ \dot\te_l}{ \tan\te_l\tan\te_{j+1}}
- \dot\te_{j+1}\right), \\
&{\dot s}_N= \Sk_\k(\tea)\prod_{m=2}^{N}\sin\te_m\left(
\frac{ \dot\tea}{ \Tk_\k(\tea) }
+\sum_{l=2}^N\frac{ \dot\te_l}{  \tan\te_{l}}\right). 
\endaligned
\tag3.6
$$
We now introduce the parametrizations (2.14), (2.17), velocities (3.5),
(3.6)  as well as the momenta (3.2) in (3.4), obtaining a  phase space
realization of the generators of  $\frak {so}_\k(N+1)$
 given in ``parallel canonical" coordinates by $(i,j=1,\dots,N)$:
$$
\aligned
&\tildep_i=\prod_{k=1}^i\! \Ck_\k(\a_k)\Ck_\k(\a_i)p_i
+\k\Sk_\k(\a_i)\sum_{s=1}^i\Sk_\k(\a_s)
\frac{\prod_{m=1}^s \Ck_\k(\a_m)}{\prod_{l=s}^i \Ck_\k(\a_l)}\,p_s ,\\
&\tildej_{ij}=\Sk_\k(\a_i)\Ck_\k(\a_j)\prod_{s=i+1}^j\!\!  \Ck_\k(\a_s)
p_j-
\frac{\Ck_\k(\a_i)\Sk_\k(\a_j)}{\prod_{k=i+1}^j\Ck_\k(\a_k)}\,p_i \\
& \qquad +\k\Sk_\k(\a_i)\Sk_\k(\a_j)\sum_{s=i+1}^j\Sk_\k(\a_s)
\frac{\prod_{m=i+1}^s  \Ck_\k(\a_m)}{\prod_{l=s}^j  \Ck_\k(\a_l)}\,p_s , 
\endaligned
\tag3.7
$$
and in geodesic polar
coordinates and momenta  $(\te,\pp)$
 by $(i,j=1,\dots,N-1)$:
$$
\aligned
&\tildep_i=\frac{\prod_{k=2}^{i+1}\sin\te_k}{\tan\te_{i+1}}\,\pp_1
+\sum_{s=2}^{i+1}
\frac{\prod_{m=s}^{i+1}\sin\te_m \,\cos\te_s \pp_s}
{\Tk_\k(\tea)\tan\te_{i+1}\prod_{l=2}^s\sin\te_l}
-\frac{\pp_{i+1}}{\Tk_\k(\tea)\prod_{l=2}^{i+1}\sin\te_l}  ,\\
&\tildep_N= \prod_{k=2}^{N}\sin\te_k \,\pp_1
+\sum_{s=2}^{N}
\frac{ \prod_{m=s}^{N}\sin\te_m \cos\te_s}
{\Tk_\k(\tea) \prod_{l=2}^s\sin\te_l}\, \pp_s  ,\\
&\tildej_{ij}= \sin\te_{i+1}\cos\te_{j+1}\!\prod_{k=i+1}^{j}\!\sin\te_k
\,\pp_{i+1} -\frac{\cos\te_{i+1}\sin\te_{j+1}}
{\prod_{l=i+1}^j\sin\te_l}\,\pp_{j+1}  \\
&\qquad\qquad +\cos\te_{i+1}\cos\te_{j+1}\sum_{s=i+1}^{j}
\frac{ \prod_{m=s}^{j}\sin\te_m \cos\te_s}
{  \prod_{l=i+1}^s\sin\te_l}\, \pp_s ,\\
&\tildej_{iN}= \sin\te_{i+1}\!\prod_{k=i+1}^{N}\!\sin\te_k \,\pp_{i+1}
+\cos\te_{i+1}\sum_{s=i+1}^{N}
\frac{ \prod_{m=s}^{N}\sin\te_m \cos\te_s}
{  \prod_{l=i+1}^s\sin\te_l}\, \pp_s  .
\endaligned
\tag3.8
$$

Then the following results follow  \cite{4}.

\proclaim{Proposition 3.1}  Both sets of  generators   (3.7) and
(3.8)  fulfil  the commutation rules (2.1) of $\frak{so}_\k(N+1)$  with
respect to the canonical Poisson bracket (1.5).  

\endproclaim

\proclaim{Proposition 3.2}  Any generator (3.7) and
(3.8)  Poisson-commutes with the kinetic energy ${\Cal T}$ (3.3).  
\endproclaim

The second statement is straightforward as   ${\Cal T}$ is related
with a phase space realization of the second-order Casimir ${\Cal  C}$ of
${so}_\k(N+1)$  through
$$
2{\Cal T}= {\tilde{\Cal  C}}=\sum_{i=1}^N \tildep^2_i+\k
\sum_{i,j=1}^N 
\tildej_{ij}^2 .
\tag3.9
$$
In fact, the geodesic motion is maximally superintegrable and its  
integrals of motion   come from any function of a phase space
realization of the Lie generators.

%%%%%%%%%%%%%%%%%%%%%%%%%%%%%%%%%%%%%%%%%%%%%%%%%%%%

\bigskip
\bigskip
\noindent
\head 4.  
Smorodinsky--Winternitz system on curved spaces\endhead

\medskip
\noindent
The problem now is to find  which potentials ${\Cal U}(q)$  
can be added to ${\Cal T}$  in such a manner that the new Hamiltonian
${\Cal H}={\Cal T}+ {\Cal U}$ preserves the maximal
superintegrability.  This   requires to add
``some" terms to ``some" functions of the generators in order
to ensure their involutivity with respect to ${\Cal H}$.

\bigskip
\medskip
\noindent
\subhead 4.1. Potential\endsubhead

\medskip
\noindent
As far as the 2D potential   (1.1) is concerned, the generalization to 
 the space
$S^N_{[\k]}$ is straightforward and reads  
$$
{\Cal U}(s(q))  = \be_0\frac { \sum_{l=1}^N s_l^2}{s_0^2}+ \sum_{i=1}^N
\frac {\be_i}{s_i^2} ,
\tag4.1
$$
 which in geodesic parallel and polar coordinates turns out to be
$$
\aligned
{\Cal U} & =  \be_0 \sum_{i=1}^N\frac{\Sk_\k^2(\a_i)}{\prod_{l=1}^i
\! \Ck_\k^2(\a_l)}+\sum_{i=1}^{N-1}\frac{\be_i}
{\Sk_\k^2(\a_i)\prod_{l=i+1}^N\! \Ck_\k^2(\a_l)}+
\frac{\be_N}{\Sk_\k^2(\a_N)}\\
 &=  \be_0
\Tk_\k^2(\tea) 
 +\frac{1}{\Sk_\k^2(\tea)}\left(\frac{\be_1}{\cos^2\te_2}
  +\sum_{i=2}^{N-1} \frac{\be_i}{ \cos^2\te_{i+1} 
\prod_{l=2}^i\sin^2\te_l  }+\frac{\be_N}{ \prod_{l=2}^N\sin^2\te_l} 
\right) .
\endaligned
\tag4.2
$$

This potential, which coincides with (1.1) for $\k=1$,  has
been interpreted  on $\Bbb S^2$ as three Higgs spherical
oscillators, whose ``centers" are placed at the three vertices
of a  sphere's octant~\cite{29} (recall that the Higgs oscillator has
two antipodal ``centers"); this interpretation can 
straightforwardly be extended to  the sphere  
$\Bbb S^N$ with
$\k=1/R^2>0$. Let $O_i$ be the points placed along the basic geodesics $l_i$ and
a quadrant apart from ${\Cal O}$ (each pair taken from ${\Cal O}, O_i$ are mutually
separated  a quadrant distance $\pi/(2\sqrt{\k})=R\,\pi/2$ on $\Bbb S^N$). If we set
$\k=1$ and denote $r,r_i$ the distances
between any generic point $Q$ and $\Cal O$, $O_i$ along the joining geodesics,  
we find 
$s_0=\cos r$, $s_i=\cos r_i$ $(i=1,\dots,N)$, and (4.2) can be rewritten as
$$
{\Cal U}=\be_0\tan^2\tea +\sum_{i=1}^N
\frac{\be_i}{\cos^2 r_i}  =\be_0\tan^2\tea +\sum_{i=1}^N \be_i\tan^2 r_i
+ \sum_{i=1}^N{\be_i} ,
\tag4.3
$$
which can  clearly be recognized under  this form as the joint potential due to a
set of $N+1$ spherical oscillators whose centers are at the $N+1$
points $\Cal O$, $O_i$. Alternatively, if $\dis_i=\pi/2-r_i$, then each
``$\beta_i$" term in (4.3) can   be described, as
usual, as the spherical ``centrifugal" barriers $\beta_i/\sin^2 \dis_i$. Under the
contraction
$\k=0$, $\Bbb S^N\to
\Bbb E^N$, the Higgs-term    gives rise to the ``flat" harmonic
oscillator $\be_0\tea^2=\be_0\sum_i\a_i^2$, while the  $N$ remaining
oscillators (whose centers would be now ``at infinity") leave the
Euclidean ``centrifugal" barriers
$\be_i/\dis_i^2 \equiv  \be_i/a_i^2$ as their imprints.

\topinsert
\epsfbox{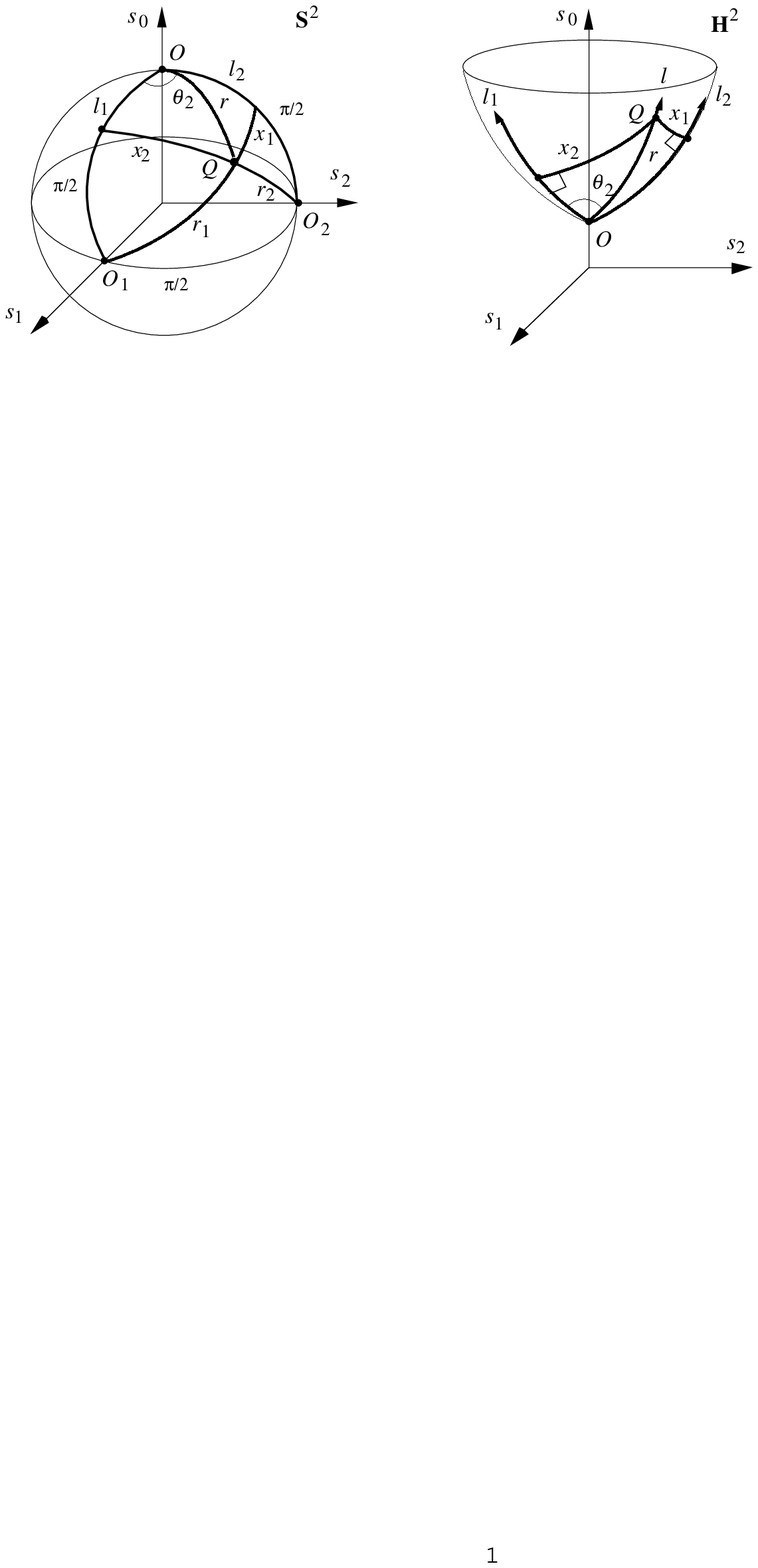}
\botcaption{Figure 2} Distances involved in   the
SW potential on $\Bbb S^2$ and $\Bbb H^2$.
\endcaption
\endinsert

On the hyperbolic space $\Bbb H^N$  the potential (4.2) can similarly be
 interpreted. Let $\dis_i$ be the
distance between the generic point $Q$ and the totally geodesic
codimension one submanifold through ${\Cal O}$ orthogonal to
the geodesic $l_i$. When  $\k=-1$ we find that $s_0=\cosh r$, $s_i=\sinh
\dis_i$ $(i=1,\dots,N)$, so the potential   can be written   as:
$$
{\Cal U}=\be_0\tanh^2\tea +\sum_{i=1}^N
\frac{\be_i}{\sinh^2 \dis_i}  =\be_0\tanh^2\tea +\sum_{i=1}^N
\frac{\be_i}{\tanh^2 \dis_i}  - \sum_{i=1}^N{\be_i} . 
\tag4.4
$$
The first term is   an ``actual" hyperbolic oscillator with center 
 at   ${\Cal O}$ (whose potential is bounded in the whole hyperbolic
space), while each of the 
$N$ remaining terms can be interpreted, as done conventionally, as some
kind of hyperbolic ``centrifugal" potentials, but also as   ``ideal"
hyperbolic oscillators, whose centers would be beyond infinity, that is, in
the exterior region of the hyperbolic space. 

%%%%%%%%%%%%%%%%%%%%%%%%%%%%%%%%%%%%%%%%%%%%%%%%%%%%

\bigskip
\medskip
\noindent
\subhead 4.2. Integrals of motion\endsubhead

\medskip
\noindent
By taking
into account the  results given in \cite{27} for  the integrals of motion
of (1.1), let us consider the
following functions,  with quadratic
dependence on momenta, given in  the ambient space
$\Bbb R^{N+1}$ by  $(i<j,\  i,j=0,1,\dots,N)$:
$$
I_{ij}=(s_i {\dot s}_j-   s_j {\dot
s}_i)^2+2\be_i\frac{s_j^2}{s_i^2}+2\be_j\frac{s_i^2}{s_j^2} .
\tag4.5
$$
Therefore we have $N(N+1)/2$
phase space functions coming from the Lie generators, which in the
geodesic parallel canonical coordinates (3.7) turn out to be
$$
\aligned
&I_{0i}
=\tildep_i^2+2\be_0\,\frac{\Sk_\k^2(\a_i)}{\prod_{l=1}^i\Ck_\k^2(\a_l)}
+2\be_i\,\frac{\prod_{l=1}^i\Ck_\k^2(\a_l)}{\Sk_\k^2(\a_i)} ,\\
&I_{ij}=\tildej_{ij}^2+2\be_i\,\frac{\Sk_\k^2(\a_j)}{
\Sk_\k^2(\a_i)\prod_{l=i+1}^j\Ck_\k^2(\a_l)}
+2\be_j\,\frac{
\Sk_\k^2(\a_i)\prod_{l=i+1}^j\Ck_\k^2(\a_l)}{\Sk_\k^2(\a_j)} .
\endaligned
\tag4.6
$$
The same quantities read in the geodesic polar canonical coordinates
(3.8):
$$
\aligned
&I_{0i}
=\tildep_i^2+2\be_0\,
\frac{\Tk^2_\k(\tea)\prod_{l=2}^{i+1}\sin^2\te_l  }{\tan^2\te_{i+1}}+
2\be_i\frac{\tan^2\te_{i+1}}{\Tk^2_\k(\tea)\prod_{l=2}^{i+1}\sin^2\te_l
},\\ &I_{0N}
=\tildep_N^2+2\be_0\,
 {\Tk^2_\k(\tea)\prod_{l=2}^{N}\sin^2\te_l } +
2\be_N\frac{ 1}{\Tk^2_\k(\tea)\prod_{l=2}^{N}\sin^2\te_l},\\
&I_{ij}
=\tildej_{ij}^2+2\be_i\,
\frac {\cos^2\te_{j+1}\prod_{l=i+1}^{j}\sin^2\te_l }{\cos^2\te_{i+1}} +
\frac{ 2\be_j \cos^2\te_{i+1} }
{\cos^2\te_{j+1}\prod_{l=i+1}^{j}\sin^2\te_l },\\ 
&I_{iN}=\tildej_{iN}^2+2\be_i\,
\frac {\prod_{l=i+1}^{N}\sin^2\te_l  }{\cos^2\te_{i+1}} + 
2\be_N\frac{ \cos^2\te_{i+1}}  {\prod_{l=i+1}^{N}\sin^2\te_l  } . 
\endaligned
\tag4.7
$$

\medskip
\medskip

A first property for these quantities is given by:
\medskip

\proclaim{Proposition 4.1} The $N(N+1)/2$ functions given by either (4.6)
or (4.7) fulfil
$ \left\{I_{ij},I_{lm}   \right\}=0$ whenever 
all four indices $i<j; l<m$, are different.

\endproclaim
\medskip

From now on we consider the
Hamiltonian ${\Cal H}={\Cal T}+{\Cal U}$ with ${\Cal T}$ and  ${\Cal
U}$ given in  (3.3) and (4.2). Then we find that:
\medskip
\medskip

\proclaim{Proposition 4.2} The $N(N+1)/2$ functions either (4.6) or (4.7)
are integrals of the motion for the Hamiltonian, that is,
$\left\{I_{ij},{\Cal H}\right\}=0$ $\forall ij$.
\endproclaim
\medskip

We remark that the
 property analogous to (3.9) is now given by
$$
2{\Cal H}= \sum_{i=1}^N I_{0i}+\k
\sum_{i,j=1}^N  I_{ij} + 2 \k\sum_{i=1}^N\be_i .
\tag4.8
$$
Notice also that when $\k=0$, the Hamiltonian expressed in parallel
coordinates and the integrals (4.6) with generators (3.7) directly  reduce
to the flat SW system characterized by (1.3) and (1.4).

%%%%%%%%%%%%%%%%%%%%%%%%%%%%%%%%%%%%%%%%%%%%%%%%%%%%

\bigskip
\medskip
\noindent
\subhead 4.3. Maximal superintegrability\endsubhead

\medskip
\noindent
The last step is firstly to identify, within the initial
set of $N(N+1)/2$ integrals of the motion together with the Hamiltonian, $N$ which
are functionally independent and in involution in order to prove the complete
integrability of ${\Cal H}$, and secondly to find out how many  
are  functionally independent  thus analyzing its superintegrability.

 Let us choose two subsets of $N-1$ integrals
$\inte^{(l)}$  and $\inte_{(l)}$     ($l=2,\dots,N$) 
obtained starting from the integrals  $I_{ij}$ associated to the
rotation generators as:
$$
 \inte^{(l)} =\sum_{i,j=1}^l
I_{ij} ,\qquad  
 \inte_{(l)} =\!\!\sum_{i,j=N-l+1}^N\!\!
I_{ij}     ,
\tag4.9
$$
which share the element
$\inte^{(N)}\equiv\inte_{(N)}$. The  generators that determine the
quadratic terms in the momenta in the first set $\inte^{(l)}$  are
associated to a chain of orthogonal subalgebras within 
${\frak h}={\frak {so}}(N)=\langle J_{ij}\rangle$ (2.3) starting
``upwards" from $\frak{so}(2)=\langle J_{12}\rangle$; likewise 
for $\inte_{(l)}$ but starting
``backwards" from $\frak{so}(2)=\langle J_{N-1\,N}\rangle$:

$$
\alignedat{19}
&\inte^{(2)}&\subset\    &\dots &\subset\  
&\inte^{(l)}&\subset\ 
&\dots &\subset\    &\inte^{(N)}
&\qquad\inte_{(N)}&\supset\    &\dots &\supset\  
&\inte_{(l)}&\supset\ 
&\dots &\supset\    &\inte_{(2)}\\
&\frak{so}(2)\ &\subset\     &\dots &\subset\  
&\frak{so}(l)&\ \subset\ 
&\dots &\subset\    &\frak{so}(N)
&\qquad\frak{so}(N)\ &\supset\     &\dots &\supset\  
&\frak{so}(l)&\ \supset\ 
&\dots &\supset\    &\frak{so}(2)
\endalignedat
$$

We remark that  the SW Hamiltonian    on $\Bbb E^N$, ${\Cal
H}\bigl|_{\k=0}$,
 can also be constructed through a coalgebra approach  \cite{5} by means
of $N$ copies of $\frak{sp}(2,\Bbb R)\simeq \frak{sl}(2,\Bbb R)$ \cite{1}. In this
flat case, each integral of motion $\inte^{(l)}\bigl|_{\k=0}$ in 
 Cartesian  coordinates {\it coincides} with an $N$-particle phase space
realization of the
$k$-th order left-coproduct of the Casimir of $\frak{sl}(2,\Bbb R)$; similarly 
for   $\inte_{(l)}\bigl|_{\k=0}$ using the right-coproduct (these are called left-
and right-integrals, respectively). Such coalgebra method
shows that the   $N$ functions
$\{  \inte^{(l)}\bigl|_{\k=0}, {\Cal H}\bigl|_{\k=0} \}$ 
are mutually   in involution (the right-set as well), and  the $2N-2$ functions 
$\{  \inte^{(l)}\bigl|_{\k=0},\inte_{(l)}\bigl|_{\k=0}, {\Cal H}\bigl|_{\k=0} \}$
are functionally independent~\cite{2}.  For the general case with arbitrary
curvature $\k$,   we have the following result.

\medskip
\medskip

\proclaim{Theorem 4.3} 
(i) The   $N$  functions $$\{
\inte^{(2)},\inte^{(3)},\dots,\inte^{(N)},\Cal H\} $$ are mutually   in
involution.
The same property holds for the second set    $$\{
\inte_{(N)},\dots,\inte_{(3)},\inte_{(2)},\Cal H\} .$$

\noindent
(ii) The $2N-1$ functions $$\{
\inte^{(2)}, \inte^{(3)},\dots,
 \inte^{(N)}\equiv \inte_{(N)},\dots,\inte_{(3)}, 
\inte_{(2)}, I_{0i},\Cal H\},$$ where $i$ is fixed, are functionally
independent, thus   
$\Cal H$ is a  maximally  superintegrable Hamiltonian. 

\endproclaim

\medskip

Notice that if $\k\ne0$, each $\inte^{(l)}$ (or 
$\inte_{(l)}$) can be seen as a smooth function on the curvature:
$$\inte^{(l)}= \inte^{(l)}\bigl|_{\k=0}+\k\,
\inte^{(l)}_1+o(\k^2) .$$ The same holds if one introduces a quantum deformation of 
the Euclidean  SW system  in terms of  a deformation parameter $z$ 
\cite{1, 2}: superintegrability is preserved and the deformed integrals
$\inte^{(l)}_z$ (or $\inte_{(l)}^z$) can be written as   power series of
$z$: $$\inte^{(l)}_z=\inte^{(l)}_z\bigl|_{z=0} +z\,\inte^{(l)}_{z,1}+o(z^2). $$ 
Then, $\inte^{(l)}\bigl|_{\k=0}\equiv \inte^{(l)}_z\bigl|_{z=0}$ are the integrals
of the SW system on $\Bbb E^N$.   This suggests some kind of duality between
quantum deformations ($z$) and curvature ($\k$)   \cite{3}.

%%%%%%%%%%%%%%%%%%%%%%%%%%%%%%%%%%%%%%%%%%%%%%%%%%%%

\bigskip
\bigskip
\noindent
\head 5. The $\bold{N=4}$ case\endhead

\medskip
\noindent
 In what follows we illustrate the general expressions obtained in the
previous sections by applying them to the case $N=4$ in terms of geodesic parallel
coordinates and momenta.   The SW Hamiltonian on 
$\Bbb S^4$,  $\Bbb E^4$ and  $\Bbb H^4$   reads
$$
\aligned
&{\Cal H}=\frac12 \left(  
\frac{p_1^2}{\Ck_\k^2(\a_2)\Ck_\k^2(\a_3)\Ck_\k^2(\a_4)}+
\frac{p_2^2}{\Ck_\k^2(\a_3)\Ck_\k^2(\a_4)}+
\frac{p_3^2}{\Ck_\k^2(\a_4)}+p_4^2\right) \\ 
&\qquad +\be_0\left(\Tk_\k^2(\a_1)+\frac{\Tk_\k^2(\a_2)}{\Ck_\k^2(\a_1)}+
\frac{\Tk_\k^2(\a_3)}{\Ck_\k^2(\a_1)\Ck_\k^2(\a_2)}+
\frac{\Tk_\k^2(\a_4)}{\Ck_\k^2(\a_1)\Ck_\k^2(\a_2)\Ck_\k^2(\a_3)}
\right)\\
&\qquad
+\frac{\be_1}{\Sk_\k^2(\a_1)\Ck_\k^2(\a_2)\Ck_\k^2(\a_3)\Ck_\k^2(\a_4)}
+\frac{\be_2}{\Sk_\k^2(\a_2)\Ck_\k^2(\a_3)\Ck_\k^2(\a_4)}\\
&\qquad
+\frac{\be_3}{\Sk_\k^2(\a_3) \Ck_\k^2(\a_4)}
+\frac{\be_4}{\Sk_\k^2(\a_4)  } .
 \endaligned
\tag5.1
$$

\newpage
 
The phase space realization (3.7) for  $\frak {so}_\k(5)$  is given by
four translations
$$
\aligned
&\tildep_1=p_1,\qquad
 \tildep_2=\Ck_\k (\a_1)p_2+\k \Sk_\k (\a_1) \Tk_\k (\a_2)p_1,\\
&\tildep_3=\Ck_\k (\a_1)\Ck_\k (\a_2)p_3+\k  
\Tk_\k (\a_3)\left( \frac{\Sk_\k(\a_1)}{\Ck_\k(\a_2)}\, p_1  
+   \Ck_\k (\a_1) \Sk_\k (\a_2) p_2\right),\\
&\tildep_4=\Ck_\k (\a_1)\Ck_\k(\a_2) \Ck_\k(\a_3) p_4 
 +\k  
\Tk_\k (\a_4)\left( \frac{\Sk_\k(\a_1)}{\Ck_\k(\a_2)\Ck_\k(\a_3)}\, p_1
\right .\\  
&\qquad\qquad \left. +\frac{\Ck_\k(\a_1)\Sk_\k(\a_2)}{\Ck_\k(\a_3)}\,
p_2   +  \Ck_\k (\a_1)\Ck_\k (\a_2) \Sk_\k (\a_3) p_3\right),\\
 \endaligned
\tag5.2
$$
together with six rotation generators:
$$
\aligned
&\tildej_{12}= \Sk_\k (\a_1)  p_2- \Ck_\k (\a_1)\Tk_\k (\a_2)p_1  ,\\
&\tildej_{13}= \Sk_\k (\a_1) \Ck_\k (\a_2) p_3-\frac{ \Ck_\k (\a_1)}{
\Ck_\k (\a_2)}  \Tk_\k (\a_3)p_1
+\k \Sk_\k (\a_1)\Sk_\k (\a_2) \Tk_\k
(\a_3)p_2,\\
&\tildej_{14}=\Sk_\k (\a_1) \Ck_\k (\a_2)\Ck_\k (\a_3) p_4-\frac{ \Ck_\k
(\a_1)}{ \Ck_\k (\a_2) \Ck_\k (\a_3)}  \Tk_\k (\a_4)p_1\\
&\qquad\qquad 
+\k \Sk_\k (\a_1) \Tk_\k
(\a_4) \left( \frac{ \Sk_\k (\a_2)}{ \Ck_\k (\a_3)} p_2 
+ \Ck_\k (\a_2)\Sk_\k (\a_3)  p_3\right),\\
&\tildej_{23}= \Sk_\k (\a_2)  p_3- \Ck_\k (\a_2)\Tk_\k (\a_3)p_2  ,\\
&\tildej_{24}= \Sk_\k (\a_2) \Ck_\k (\a_3) p_4-\frac{ \Ck_\k (\a_2)}{
\Ck_\k (\a_3)}  \Tk_\k (\a_4)p_2
+\k \Sk_\k (\a_2)\Sk_\k (\a_3) \Tk_\k
(\a_4)p_3,\\
&\tildej_{34}= \Sk_\k (\a_3)  p_4- \Ck_\k (\a_3)\Tk_\k (\a_4)p_3  .
 \endaligned
\tag5.3
$$
From  these generators we construct ten integrals of motion (4.6):
$$
\alignedat{9}
&\tildep_{1}&\ \ &\tildep_{2} &\ \ &\tildep_{3} &\ \ &\tildep_{4} &\quad & 
&\quad &I_{01}&\ \ &I_{02} &\ \ &I_{03} &\ \ &I_{04} \cr 
\noalign{\vskip -0.3cm}
&\multispan7\hrulefill&&\omit&&\multispan7\hrulefill\\ 
\noalign{\vskip -0.1cm}
& &\ \ &\tildej_{12} &\ \ &\tildej_{13} &\ \ &\tildej_{14} &\qquad
&\Longrightarrow  &\qquad & &\ \ &I_{12} &\ \ &I_{13} &\ \ &I_{14} \\ 
& &\ \ &  &\ \ &\tildej_{23} &\ \ &\tildej_{24} &\quad & 
&\quad & &\ \ &  &\ \ &I_{23} &\ \ &I_{24} \\ 
& &\ \ &  &\ \ & &\ \ &\tildej_{34} &\quad & 
&\quad & &\ \ &  &\ \ &  &\ \ &I_{34} 
\endalignedat
$$
Hence from (5.2) we obtain four ``translation-like" integrals  given by
$$
\aligned
&I_{01}=\tildep^2_1+2\be_0 \Tk^2_\k (\a_1)+2\be_1
\frac{1}{\Tk^2_\k(\a_1)},\\ &I_{02}=\tildep^2_2 +
2\be_0\frac{\Tk^2_\k(\a_2)}{\Ck^2_\k(\a_1)}+
 2\be_2\frac{\Ck^2_\k(\a_1)}{\Tk^2_\k(\a_2)},\\
&I_{03}=\tildep^2_3  +
2\be_0\frac{\Tk^2_\k(\a_3)}{\Ck^2_\k(\a_1)\Ck^2_\k(\a_2)}+
 2\be_3\frac{\Ck^2_\k(\a_1)\Ck^2_\k(\a_2)}{\Tk^2_\k(\a_3)},\\
&I_{04}=\tildep^2_4  +
2\be_0\frac{\Tk^2_\k(\a_4)}{\Ck^2_\k(\a_1)\Ck^2_\k(\a_2)\Ck^2_\k(\a_3)}+
2\be_4\frac{\Ck^2_\k(\a_1)\Ck^2_\k(\a_2)\Ck^2_\k(\a_3)}{\Tk^2_\k(\a_4)} ,
 \endaligned
\tag5.4
$$
together with six ``rotation-like" ones  coming from (5.3):
$$
\aligned
&I_{12}=\tildej_{12}^2+2\be_1
\frac{ \Tk^2_\k (\a_2)}{\Sk^2_\k(\a_1)}
+2\be_2 \frac{\Sk^2_\k(\a_1)}{ \Tk^2_\k (\a_2)},\\
&I_{13}=\tildej_{13}^2+2\be_1
\frac{ \Tk^2_\k (\a_3)}{\Sk^2_\k(\a_1)\Ck^2_\k(\a_2)}
+2\be_3 \frac{\Sk^2_\k(\a_1)\Ck^2_\k(\a_2)}{ \Tk^2_\k (\a_3)},\\
&I_{14}=\tildej_{14}^2+2\be_1
\frac{ \Tk^2_\k (\a_4)}{\Sk^2_\k(\a_1) \Ck^2_\k(\a_2)\Ck^2_\k(\a_3)}
+2\be_4 \frac{\Sk^2_\k(\a_1)\Ck^2_\k(\a_2)\Ck^2_\k(\a_3)}{ \Tk^2_\k
(\a_4)},\\
&I_{23}=\tildej_{23}^2+2\be_2
\frac{ \Tk^2_\k (\a_3)}{\Sk^2_\k(\a_2)}
+2\be_3 \frac{\Sk^2_\k(\a_2)}{ \Tk^2_\k (\a_3)},\\
&I_{24}=\tildej_{24}^2+2\be_2
\frac{ \Tk^2_\k (\a_4)}{\Sk^2_\k(\a_2)\Ck^2_\k(\a_3)}
+2\be_4 \frac{\Sk^2_\k(\a_2)\Ck^2_\k(\a_3)}{ \Tk^2_\k (\a_4)},\\
&I_{34}=\tildej_{34}^2+2\be_3
\frac{ \Tk^2_\k (\a_4)}{\Sk^2_\k(\a_3)}
+2\be_4 \frac{\Sk^2_\k(\a_3)}{ \Tk^2_\k (\a_4)} .
 \endaligned
\tag5.5
$$

Thus, in this case, within the subset of ``rotation-like" integrals, we find
 three ``upwards-integrals" $
\inte^{(2)},\inte^{(3)},
\inte^{(4)}$, associated to $\frak{so}(2)\subset \frak{so}(3)\subset
\frak{so}(4)$, that determine a completely integrable Hamiltonian: 
$$
\matrix \inte^{(2)}= I_{12}\\ \\ \\ \endmatrix  \qquad 
\matrix\inte^{(3)}= I_{12}+I_{13}\\
\qquad\qquad\quad  +I_{23}\\  \\ \endmatrix  \qquad
\matrix \inte^{(4)}=\inte_{(4)}=I_{12}+I_{13}+I_{14}\\
\qquad\qquad\qquad\qquad\   +I_{23}+I_{24}\\
\qquad\qquad\qquad\qquad\qquad\   +I_{34}\endmatrix
$$
These three functions together with the two  ``backwards-integrals"  
$ \inte_{(2)}, \inte_{(3)}$, associated
to $\frak{so}(2)\subset \frak{so}(3)$, and one additional
``translation-like" one, say
$I_{01}$,   characterize   a maximally superintegrable
Hamiltonian (5.1):
$$
\matrix  \inte_{(2)}= I_{34}\\ \\ \endmatrix  \qquad 
\matrix  \inte_{(3)}= I_{23}+I_{24}\\
\qquad\qquad\quad   +I_{34}\\  \endmatrix 
\qquad 
\matrix  I_{01}\\ \\ \endmatrix 
$$

To end with, we would like to point out that a similar algebraic
construction may likely be applied to the curved version of the
  ``Kepler--Coulomb" superintegrable family (1.2). Furthermore, the
consideration of a second contraction parameter, say  $\k_2$, that enables
to take into account indefinite metrics of Lorentzian signature 
\cite{14, 15}, would allow one to  obtain superintegrable systems on
different spacetimes. The application of these maximally superintegrable
systems  in quantum mechanics also deserves a further study.

%%%%%%%%%%%%%%%%%%%%%%%%%%%%%%%%%%%%%%%%%%%%%%%%%%%%

\bigskip
\bigskip
 \noindent
\head  Acknowledgments\endhead

\medskip
\noindent
 This work was partially supported  by the Ministerio de Ciencia y
Tecnolog\1a, Spain (Projects BFM2000-1055 and  BFM2002-3773).
A.B.\ and F.J.H.\ are also grateful to G.S. Pogosyan for helpful
discussions and to the CRM for hospitality.

\newpage

%%%%%%%%%%%%%%%%%%%%%%%%%%%%%%%%%%%%%%%%%%%%%%%%%%%%

\bigskip
\bigskip

 \noindent
\head  References\endhead

 \medskip

 \eightpoint

\ref\key{1}
\by  A. Ballesteros and  F. J. Herranz
\paper Integrable deformations of oscillator chains from quantum algebras
\jour J. Phys.  A: Math. Gen.
\vol 32
\yr 1999
\pages 8851--8862
\endref

\ref\key{2}
\by  A. Ballesteros,  F. J. Herranz, F. Musso, and O. Ragnisco
\paper Superintegrable deformations of the Smorodinsky--Winternitz Hamiltonian
\jour in  this volume
\vol 
\yr  
\pages  
\endref

\ref\key{3}
\by  A. Ballesteros, F. J. Herranz, M. A. del Olmo, and M. Santander 
\paper Classical deformations, Poisson--Lie contractions, and quantization
of dual Lie bialgebras
\jour J. Math. Phys.
\vol 36
\yr 1995
\pages 631--640
\endref

\ref\key{4}
\by  A. Ballesteros, F. J. Herranz,  M. Santander, and T. Sanz-Gil 
\paper  Maximal superintegrability on  N-dimensional curved spaces
\jour  J. Phys.  A: Math. Gen.
\vol  36
\yr  2003
\pages  L93--L99
\endref

\ref\key{5}
\by   A. Ballesteros and O. Ragnisco
\paper A systematic construction of integrable Hamiltonians from
coalgebras
\jour   J. Phys.  A: Math. Gen.
\vol 31
\yr 1998
\pages 3791--3813
\endref

\ref\key{6}
\by  N. W.  Evans 
\paper  Superintegrability in classical mechanics
\jour   Phys. Rev. A
\vol 41
\yr 1990
\pages 5666--5676
\endref

\ref\key{7}
\by  N. W.  Evans 
\paper  Superintegrability of the  Winternitz system
\jour    Phys. Lett. A
\vol 147
\yr 1990
\pages 483--486
\endref

 \ref\key{8}
\by  N. W.  Evans 
\paper  Group theory of the Smorodinsky--Winternitz system
\jour    J. Math. Phys.
\vol 32
\yr  1991
\pages 3369--3375
\endref

 \ref\key{9}
\by  J. Fris, V. Mandrosov, Ya A. Smorodinsky, M. Uhlir, and P. Winternitz 
\paper  On higher symmetries in quantum mechanics
\jour    Phys. Lett.
\vol 16
\yr 1965
\pages 354--356
\endref

 \ref\key{10}
\by   C. Grosche, G. S. Pogosyan,  and A. N. Sissakian 
\paper  Path integral discussion for Smorodinsky--Winternitz potentials 
1. 2-dimensional and    3-dimensional Euclidean space
\jour   Fortschr. Phys. 
\vol 43
\yr 1995
\pages 453--521
\endref

\ref\key{11}
\by   C. Grosche, G. S. Pogosyan,  and A. N. Sissakian 
\paper  Path integral discussion for Smorodinsky--Winternitz  potentials
2. The 2-dimensional and    3-dimensional sphere
\jour   Fortschr. Phys. 
\vol 43
\yr 1995
\pages 523--563
\endref

 \ref\key{12}
\by   C. Grosche, G. S. Pogosyan,  and A. N. Sissakian 
\paper  Path integral approach for superintegrable potentials on the
three-dimensional hyperboloid
\jour   Phys.  Part. Nuclei
\vol 28
\yr 1997
\pages 486--519
\endref

\ref
\key{13}
\by         S. Helgason
\book      Differential Geometry and Symmetric Spaces
\publ Academic Press
\publaddr New York
\yr         1962
\endref

\ref\key{14}
\by  F. J. Herranz, R. Ortega, and M. Santander
\paper Trigonometry of spacetimes: a new self-dual approach to a
      curvature/signature (in)dependent trigonometry
\jour J. Phys.  A: Math. Gen.
\vol 33
\yr 2000
\pages 4525--4551
\endref

\ref\key{15}
\by  F. J. Herranz and M. Santander
\paper Conformal   symmetries of spacetimes
\jour J. Phys.  A: Math. Gen.
\vol 35
\yr 2002
\pages 6601--6618
\endref

\ref\key{16}
\by  P. W. Higgs
\paper Dynamical symmetries in a spherical geometry I
\jour J. Phys.  A: Math. Gen.
\vol 12
\yr 1979
\pages 309--323
\endref

 \ref\key{17}
\by   A. A. Izmest'ev,  G. S. Pogosyan, A. N. Sissakian,  and  P.
Winternitz
\paper  Contractions of Lie algebras and separation of variables. The
$n$-dimensional sphere
\jour   J. Math. Phys.
\vol 40
\yr  1999
\pages 1549--1573
\endref

\ref\key{18}
\by  E. G. Kalnins,   J. M. Kress,  G. S.  Pogosyan, and  W. Miller 
\paper  Completeness of superintegrability in two-dimensional
constant-curvature spaces
\jour   J. Phys. A: Math. Gen.
\vol 34
\yr 2001
\pages 4705--4720
\endref

 \ref\key{19}
\by   E. G. Kalnins, W. Miller, Ye M. Hakobyan, and G. S. Pogosyan
\paper Superintegrability   on the two-dimensional  
 hyperboloid II
\jour J. Math. Phys.  
\vol 40
\yr 1999
\pages 2291--2306
\endref

 \ref\key{20}
\by   E. G. Kalnins, W. Miller, and G. S. Pogosyan
\paper Superintegrability   of the two-dimensional  
 hyperboloid
\jour J. Math. Phys.  
\vol 38
\yr 1997
\pages 5416--5433
\endref

\ref\key{21}
\by  E. G. Kalnins,   W. Miller, and  G. S.  Pogosyan
\paper  Completeness of multiseparable superintegrability on the complex
2-sphere
\jour   J. Phys. A: Math. Gen.
\vol 33
\yr 2000
\pages 6791--6806
\endref

\ref\key{22}
\by   E. G. Kalnins, W. Miller, and G. S. Pogosyan
\paper  Coulomb--oscillator duality in spaces of constant curvature
\jour   J. Math. Phys.  
\vol 41
\yr 2000
\pages 2629--2657
\endref

\ref\key{23}
\by   E. G. Kalnins, W. Miller, and G. S. Pogosyan
\paper The Coulomb--oscillator relation on $n$-dimensional spheres and
hyperboloids
\jour   Phys.  Atom. Nucl.
\vol 65
\yr 2002
\pages 1086--1094
\endref

\ref\key{24}
\by   E. G. Kalnins,  G. S. Pogosyan, and W. Miller 
\paper Completeness of multiseparable superintegrability in two dimensions
\jour   Phys.  Atom. Nucl.
\vol 65
\yr 2002
\pages 1033--1035
\endref

\ref\key{25}
\by  E. G. Kalnins, G. C. Williams, W. Miller, and  G. S.  Pogosyan
\paper  Superintegrability in three-dimensional Euclidean space
\jour   J. Math. Phys. 
\vol 40
\yr 1999
\pages 708--725
\endref

  \ref\key{26}
\by  H. I. Leemon  
\paper Dynamical symmetries in a spherical geometry II
\jour J. Phys.  A: Math. Gen.
\vol 12
\yr 1979
\pages 489--501
\endref

\ref\key{27}
\by  M. F. Ra\~nada and M. Santander 
\paper Superintegrable systems on the two-dimensional sphere $S^2$ and
the hyperbolic plane $H^2$
\jour J. Math. Phys.  
\vol 40
\yr 1999
\pages 5026--5057
\endref

 \ref\key{28}
\by M. F. Ra\~nada  and  M. Santander 
\paper  On harmonic oscillators on the two-dimensional sphere $S^2$ and
the hyperbolic plane $H^2$
\jour   J. Math. Phys.
\vol 43
\yr  2002
\pages 431--451
\endref

 \ref\key{29}
\by   M. F. Ra\~nada, M. Santander,  and T. Sanz-Gil 
\paper   Superintegrable potentials and the superposition of Higgs
oscillators on 
  the sphere $S^2$
\publ Banach Center Publications
\publaddr Warszawa
\yr  2002
\pages   to be published
\endref

   \ref\key{30}
\by E. Schr\"odinger 
\paper  A method of determining quantum mechanical eigenvalues and
eigenfunctions
\jour    Proc. R. Ir. Acad. A
\vol 46
\yr  1940
\pages 9--16
\endref

\enddocument